\documentclass[pra,superscriptaddress,showpacs]{revtex4}
\usepackage{graphicx}
\begin{document}
\title{Spin swap gate in the presence of qubit inhomogeneity in a double
quantum dot}
\author{Xuedong Hu}
\affiliation{Condensed Matter Theory Center, Department of Physics,
University of Maryland, College Park, MD 20742-4111}
\affiliation{Department of Physics, University at Buffalo, SUNY,
 Buffalo,
NY 14260-1500}
\altaffiliation[Permanent address.]{}
\author{S. Das Sarma}
\affiliation{Condensed Matter Theory Center, Department of Physics,
University of Maryland, College Park, MD 20742-4111}
\date{\today}
\begin{abstract}
We study theoretically the effects of qubit inhomogeneity on the quantum logic
gate of qubit swap, which is an integral part of the operations of a quantum
computer.  Our focus here is to construct a robust pulse sequence for swap
operation in the simultaneous presence of Zeeman
inhomogeneity for quantum dot trapped electron spins and the finite-time
ramp-up of exchange coupling in a double dot.  We first present a
geometric explanation of spin swap operation, mapping the two-qubit
operation onto a single-qubit rotation.  We then show that in this
geometric picture a square-pulse-sequence can be easily designed to perform
swap in the presence of Zeeman inhomogeneity.  Finally, we investigate how
finite ramp-up times for the exchange coupling $J$ negatively affect the
performance of the swap gate sequence, and show how to correct the problems
numerically.
\end{abstract}

\maketitle

The study of quantum information processing has
attracted a great deal of attention in recent years because of the
potential advantages provided by quantum mechanical principles such as
superposition and entanglement \cite{Reviews}.  
Among the many proposed quantum computer (QC) architectures, solid-state
schemes such as those based on electron \cite{LD,Vrijen} and nuclear
spins \cite{Kane} and those based on superconducting circuits
\cite{Schon} are widely regarded as the best candidates in providing
scalable systems.  However, solid-state architectures also have their own
shortcomings.  For example, in supeconducting-circuit-based QC schemes,
offset charge noise due to jumps of trapped charges has been an important
limiting factor \cite{Nakamura,charge_echo,Saclay};
In semiconductors such as silicon,
complexities in their band structures can potentially
lead to significant difficulties
in qubit manipulations \cite{BHD}.  The often small energy scales in solids
lead to slower coherent operations and harder initializations.
Furthermore, in many solid-state QC schemes it is almost impossible to
have completely identical qubits.  As potential and problems coexist in
proposed solid-state QC architectures, theoretical explorations are sorely
needed to
provide effective forewarnings and guidelines.  In particular,
it is important to analyze theoretically various possible sources of
errors (arising, for example, from
imperfection, inhomogeneity, decoherence, nonadiabaticity, and in
general from deviations from ideal architectures assumed in the QC
proposals which must invariably be present in real solid-state systems)
in these solid-state QC schemes.  For example,
In the spin-based quantum dot quantum computer (QDQC) architecture 
\cite{LD,BLD,HD,Revs},
where trapped electron spins are the quantum bits (qubits),
quantum dots provide the
tags and the environment for the individual qubits.  Each
quantum dot is generally slightly different in size,
geometry, confinement potential depth, {\it g} factors, etc.  Some of these
differences can be accounted for straightforwardly by system calibration,
while others, such as inhomogeneity in the electron spin
Zeeman splitting, have to be treated more carefully.

In this paper we study how to overcome the problems arising from
qubit inhomogeneity and gate imperfections in the spin-based QDQC,
particularly focusing on the swap
operation.  It is important to emphasize here that in realistic QC
architectures, two-qubit operations include not only the entangling
operations such as controlled-NOT, but also auxilliary operations such as
swap, which are crucial components for the effective manipulation of a QC
(the function of swap in QDQC is to move non-neighboring qubits together
and apart through a quantum dot array during entangling operations).
Furthermore, controlled-NOT operation in QDQC is built upon the 
square-root-of-swap operations.  Therefore, imperfections that affect
swap will also affect controlled-NOT in general.  For instance,
previously we have shown that swap operation \cite{LD} cannot be done
precisely in one step if there is
inhomogeneity in the electron Zeeman splitting \cite{HSD}, because the
inhomogeneity breaks the symmetry of the two-spin system.
Here we first present a
geometric interpretation of two-spin swap operation in terms of single
spin rotation, and design a square pulse sequence to perform swap in the
presence of Zeeman inhomogeneity.  We then discuss the effects of qubit
inhomogeneity and imperfect control of the exchange coupling on the swap
gate sequence, and present its numerically corrected version.

Under the condition that the double quantum dot low-energy dynamics can be
described by Heisenberg exchange Hamiltonian \cite{HD,Gate}, the
two-electron spin Hamiltonian can be written as a sum of the exchange term
and the Zeeman splittings:
\begin{equation}
H = J {\bf S}_1 \cdot {\bf S}_2 + \gamma_1 S_{1z} + \gamma_2 S_{2z}\,,
\label{H_0}
\end{equation}
where $J$ gives the strength of exchange coupling and is a function of
quantum dot size, geometry, and confinement.  $\gamma_1$ and
$\gamma_2$ are the Zeeman splittings of the two spins and are functions
of local {\it g} factors and magnetic fields.  If we express this Hamiltonian
on the two-spin basis $|\uparrow\uparrow\rangle$,
$|\uparrow\downarrow\rangle$, $|\downarrow\uparrow\rangle$, and
$|\downarrow\downarrow\rangle$, and let $\delta=\gamma_1-\gamma_2$ be the
Zeeman inhomogeneity and $\gamma = (\gamma_1 + \gamma_2)/2$ the average
Zeeman coupling, we obtain
\begin{equation}
H = \left(
\begin{array}{cccc}
2\gamma & 0 & 0 & 0  \\
0 & \delta & J  & 0 \\
0 & J & -\delta & 0 \\
0 & 0 & 0 & -2\gamma
\end{array}
\right)\,,
\end{equation}
Notice that {\it within} this Hamiltonian the two unpolarized states
$|\uparrow\downarrow\rangle$ and $|\downarrow\uparrow\rangle$ are decoupled
from the two polarized states $|\uparrow\uparrow\rangle$ and
$|\downarrow\downarrow\rangle$, thus their dynamics can be described
separately.

To understand how swap works, we first write down the spin states before
and after a swap:
\begin{eqnarray}
\left(\begin{array}{c}
\alpha_1 \\
\alpha_2
\end{array}
\right) \otimes
\left(\begin{array}{c}
\beta_1 \\
\beta_2
\end{array}
\right) & = & \alpha_1\beta_1 |\uparrow\uparrow\rangle
+ \alpha_2\beta_2|\downarrow\downarrow\rangle
+ \alpha_1\beta_2|\uparrow\downarrow\rangle
+ \alpha_2\beta_1|\downarrow\uparrow\rangle \nonumber \\
& = & \alpha_1\beta_1 T_\uparrow + \alpha_2\beta_2 T_\downarrow
+ \frac{\alpha_1\beta_2 + \alpha_2\beta_1}{\sqrt{2}} T_0
+ \frac{\alpha_1\beta_2 - \alpha_2\beta_1}{\sqrt{2}} S \,.\nonumber \\
\left(\begin{array}{c}
\beta_1 \\
\beta_2
\end{array}
\right) \otimes
\left(\begin{array}{c}
\alpha_1 \\
\alpha_2
\end{array}
\right) & = & \alpha_1\beta_1 |\uparrow\uparrow\rangle
+ \alpha_2\beta_2|\downarrow\downarrow\rangle
+ \alpha_2\beta_1|\uparrow\downarrow\rangle
+ \alpha_1\beta_2|\downarrow\uparrow\rangle \nonumber \\
& = & \alpha_1\beta_1 T_\uparrow + \alpha_2\beta_2 T_\downarrow
+ \frac{\alpha_1\beta_2 + \alpha_2\beta_1}{\sqrt{2}} T_0
- \frac{\alpha_1\beta_2 - \alpha_2\beta_1}{\sqrt{2}} S \,,\nonumber
\end{eqnarray}
where $T_\uparrow = |\uparrow\uparrow\rangle$, $T_\downarrow =
|\downarrow\downarrow\rangle$, and $T_0 = (|\uparrow\downarrow\rangle +
|\downarrow\uparrow\rangle)/\sqrt{2}$ are the triplet states, while $S =
(|\uparrow\downarrow\rangle - |\downarrow\uparrow\rangle)/\sqrt{2}$ is the
two-spin singlet state.  From the above expressions, swap is achieved by
switching the coefficients of the unpolarized $|\uparrow\downarrow\rangle$
and $|\downarrow\uparrow\rangle$ states, or equivalently, changing the
coefficient of the singlet component by a $\pi$ phase shift relative to
the triplet states.  The phase shift can be easily obtained by a Heisenberg
exchange Hamiltonian $H=J{\bf S}_1 \cdot {\bf S}_2$, since singlet and
triplet states are the eigenstates of the exchange Hamiltonian 
(split by $4J$) so that its
only effect is to introduce dynamical phases to each basis state.

If a uniform magnetic field is present ($\gamma_1 = \gamma_2 = \gamma$),
the triplet states are split by the
Zeeman coupling $\gamma$.  When the exchange Hamiltonian is applied to the
two-spin system together with the uniform magnetic field, swap can still
be performed, with an additional phase: instead of
an exact swap, now the final states take on the form
\begin{equation}
\left(\begin{array}{c}
\alpha_1 \\
\alpha_2
\end{array}
\right) \otimes
\left(\begin{array}{c}
\beta_1 \\
\beta_2
\end{array}
\right) \rightarrow
\left(\begin{array}{c}
\beta_1 e^{i\gamma t/\hbar} \\
\beta_2 e^{-i\gamma t/\hbar}
\end{array}
\right) \otimes
\left(\begin{array}{c}
\alpha_1 e^{i\gamma t/\hbar} \\
\alpha_2 e^{-i\gamma t/\hbar}
\end{array}
\right) \,.
\end{equation}
In other words, different spin states acquire different phases
depending on their Zeeman energies \cite{HSD}.  Thus the additional phases
here come only from the polarized states.

When there is Zeeman inhomogeneity between the two quantum dots, the
spin dynamics is more complicated since the singlet and
unpolarized triplet states are coupled by the inhomogeneous Zeeman
splitting.  However, as we mentioned before, within Hamiltonian (\ref{H_0})
the dynamics of the polarized and unpolarized two-spin states are
separated.  With the polarized states still eigenstates of Hamiltonian
(\ref{H_0}), we need to focus on only the unpolarized states
$|\uparrow\downarrow\rangle$ and $|\downarrow\uparrow\rangle$.  Since these
states are not coupled to the polarized states, the two-state subspace
they span can be treated as an effective spin-$1/2$ system with an
effective Hamiltonian \cite{note}
\begin{equation}
H_e = \left(
\begin{array}{cc}
\delta & J  \\
J & -\delta
\end{array}
\right) = J \sigma_x + \delta \sigma_z = J_n \hat{n}\cdot
\mbox{\boldmath$\sigma$}\,,
\label{H_e}
\end{equation}
where $J_n = \sqrt{J^2 + \delta^2}$ and $\hat{n} = (J,0,\delta)/J_n$.
Within the picture of this effective spin-$1/2$ system, there is a simple
geometric explanation to the two-spin swap operation.  Hamiltonian
(\ref{H_e}) is a rotation on the Bloch sphere of the effective spin-$1/2$
system around the axis given by $\hat{n}$, with the rotation angle being
determined by the duration and strength of this
Hamiltonian. In the absence of inhomogeneity ($\delta=0$), the rotation is
around the $x$ axis.  Thus a $\pi$ rotation for this effective spin-$1/2$
system would correspond preceisely to
a ``swap'' in the original two-qubit system: 
$|\uparrow\downarrow\rangle \rightarrow
|\downarrow\uparrow\rangle$ and $|\downarrow\uparrow\rangle \rightarrow
|\uparrow\downarrow\rangle$.  

\begin{figure}
\includegraphics[width=3.6in]
{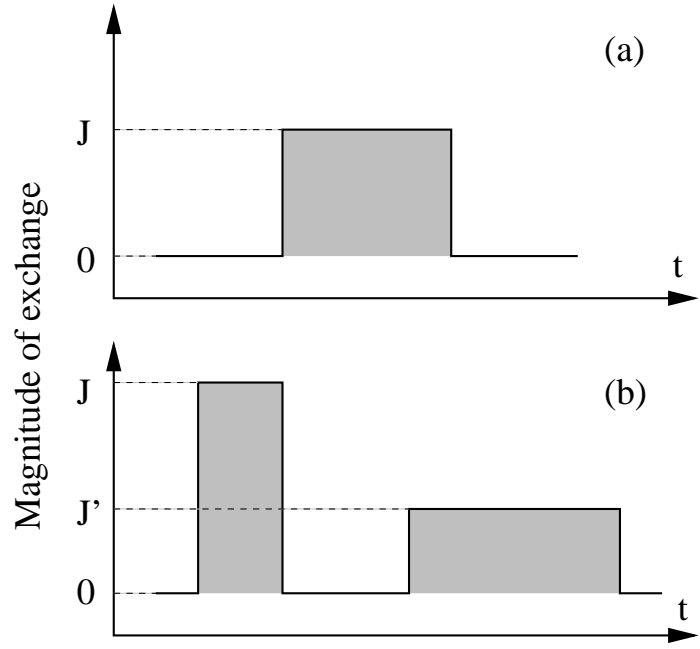}
\protect\caption[square pulse sequence for swap]
{\sloppy{Square pulse sequences for a swap operation.  (a) In the absence of
Zeeman inhomogeneity, swap can be achieved by a $\pi$ pulse of the exchange
interaction; (b) In the presence of Zeeman inhomogeneity $\delta$, swap can
be achieved by two $\pi$ pulses of the exchange interaction with magnitudes
$J$ and $J^{\prime}=\delta^2/J$, and a $\pi$ pulse for free evolution with
$\delta$ in between.
}}
\label{fig:pulses}
\end{figure}

In the presence of Zeeman inhomogeneity, the rotational axis $\hat{n}$
deviates away from the $x$ axis.  Now starting from the north pole of the
Bloch sphere, the state will not be able to reach the south pole by one
rotation around a fixed axis (corresponding to a square pulse of exchange
coupling in the presence of inhomogeneity).  
Thus, Zeeman inhomogeneity makes the exact swap impossible by a
single pulse of $J$ (with a fixed sign) \cite{HSD}. However, swap can still
be performed if the Zeeman inhomogeneity is known (for example, if it is
due to engineered g-factor). Essentially, we can adjust the rotational axis
by changing the magnitude of the exchange coupling $J$.  For
example, we can switch on a square pulse of exchange with magnitude $J$ for
a $\pi$ rotation, then turn off the exchange and let the system undergo
a $\pi$ rotation around the $z$ axis (driven by the
inhomogeneous Zeeman splitting),
then switch on a second square pulse of exchange with
magnitude $J^\prime = \delta^2/J$ for a $\pi$ rotation.  If we define
$R(\hat{n},\theta)$ to be a $\theta$-rotation around the $\hat{n}$
direction, the pulse sequence for swap would be:
\begin{equation}
U_{\rm swap} = R(\hat{n},\pi) \,R(\hat{z},\pi) \,R(\hat{n}^{\prime},\pi) \,,
\label{eq:swap}
\end{equation}
where the rotational axis are $\hat{n}=(J,0,\delta)/\sqrt{J^2 + \delta^2}$,
$\hat{z}=(0,0,1)$, and 
$\hat{n}^{\prime}=(J^{\prime},0,\delta)/\sqrt{J^{\prime \, 2} + \delta^2}$.
The end result of this pulse sequence
would be a swap for the two spins:
$|\uparrow \downarrow \rangle \leftrightarrow |\downarrow
\uparrow \rangle$.  Figure~\ref{fig:pulses} shows a schematic comparison of
the square pulse sequence in the absence and presence of Zeeman inhomogeneity.
If in the above pulse sequence the exchange couplings $J$ or $J^{\prime}$ is
unphysically large, we can always use smaller exchange couplings instead, but
with a longer pulse sequence.  For example, if $J$ is the maximal exchange
while $J^{\prime}$ is even bigger (which implies a large Zeeman inhomogeneity 
$\delta$), we can keep performing $R(\hat{n},\pi) \,R(\hat{z},\pi)$ until at
last a $R(\hat{n}^{\prime},\pi)$ with $J^{\prime}<J$ can be used to complete
the sequence ($J^{\prime}$ is determined uniquely by $J$ and $\delta$):
\begin{equation}
U_{\rm swap} = R(\hat{n},\pi) \,R(\hat{z},\pi) \,R(\hat{n},\pi)
\,R(\hat{z},\pi) \cdots
\,R(\hat{n}^{\prime},\pi) \,.
\label{eq:swap_multi}
\end{equation}
Take a simple example of $J/\delta = \tan(\pi/8)=\sqrt{2}-1$ with $J$ being
the largest possible physical value, the pulse sequence for swap is
\begin{equation}
U_{\rm swap} = R(\hat{n},\pi) \,R(\hat{z},\pi) \,R(\hat{n},\pi)
\,R(\hat{z},\pi) \,R(\hat{n},\pi) \,R(\hat{z},\pi)
\,R(\hat{n},\pi) \,,
\end{equation}
which invokes the exchange coupling $4$ times.  This simply shows that
exchange becomes less efficient in performing swap operation if Zeeman
inhomogeneity is large.  An
alternative interaction might have to be used instead (such as optically
assisted spin flip).

Using the geometric picture for swap, it is apparent that the presence
of inhomogeneous Zeeman splitting introduces additional complexities to
swap operation.  For instance, in the absence of Zeeman inhomogeneity, it
does not matter whether the exchange coupling $J$ is switched on suddenly
or gradually as the rotational axis is always fixed along $x$; while in the
presence of Zeeman inhomogeneity, swap
operation cannot be done for two spins (or, flip cannot be achieved for
the effective spin-1/2) {\it in one shot}.  The introduction of a finite pulse
rise/fall time aggravates this problem as the orientation of
the rotational axis of the effective spin-1/2 system becomes time-dependent
(a gradually switched-on exchange coupling
means that the rotational axis is time-dependent for the switch-on period,
which immediately leads to incomplete and/or imprecise rotations).  To
quantify this difference we perform a numerical calculation and show that
the shift in rotational axis reduces the efficiency of
exchange coupling in performing swap operations.  We then numerically 
search for the appropriate exchange coupling to perform swap operation
under these non-ideal conditions.

\begin{figure}
\includegraphics[width=3.6in]
{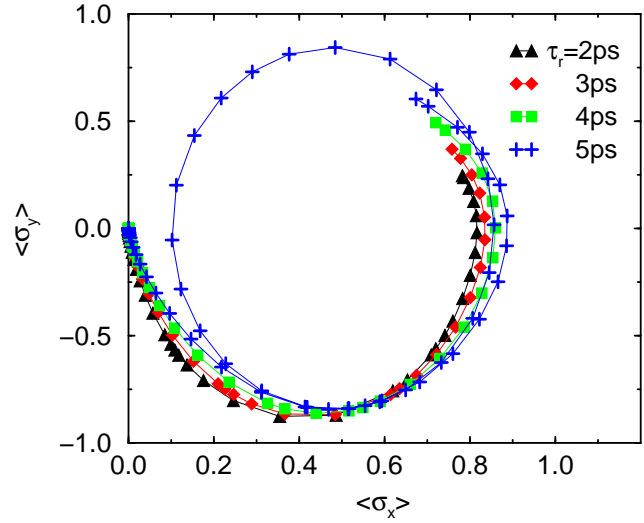}
\protect\caption[sample trajectories]
{\sloppy{Sample trajectories of rotations on the Bloch sphere of the
effective spin-$1/2$ system performed by single exchange pulses and
projected onto the $xy$ plane.
}}
\label{fig:traj}
\end{figure}

In the following we use the square pulse case (sharp rise and fall of the
exchange
coupling) as a benchmark to measure the degradation of the rotation
by the exchange pulses with finite rise/fall times and introduce an
effective infidelity $\Delta \sigma_z = \left.\sigma_z \right|_{\rm opt} -
\sigma_0$, where $\sigma_0$ is the smallest $\sigma_z$ reached by a square
exchange pulse after a $\pi$ rotation starting from the north pole of the
Bloch sphere, while $\left.\sigma_z \right|_{\rm opt}$ is the smallest
$\sigma_z$ reached by an exchange pulse with finite $\tau_r$ and optimized
pulse duration (not necessarily a $\pi$ pulse, which is ill-defined for
a time-dependent axis anyway).
Fig.~\ref{fig:traj} gives several trajectories for optimized rotation
on the Bloch
sphere projected onto the $xy$ plane.  Each trajectory represents a
case with a chosen pulse rise/fall time and an optimized pulse
duration so that the final destination point has the smallest
$\langle \sigma_z \rangle$ component (closest to the $\langle \sigma_z
\rangle = -1$ south pole point.  Recall that a rotation from the north
pole to the south pole and vice versa corresponds to an exact swap:
$|\uparrow\downarrow\rangle \leftrightarrow |\downarrow\uparrow\rangle$).  
All the curves share a common
exchange coupling $J = 0.2$meV and an inhomogeneity of $\delta=0.1$meV.
One interesting feature here is that when the pulse rise/fall time is
sufficiently long (5ps in this case), the system has to undergo
approximately one and a half full rotation (instead of one half full
rotation, or the $\pi$-pulse) in order to reach the smallest $\langle
\sigma_z \rangle$.  In addition, notice that these optimal
rotations are generally not $\pi$-rotations anymore.

\begin{figure}
\includegraphics[width=3.6in]
{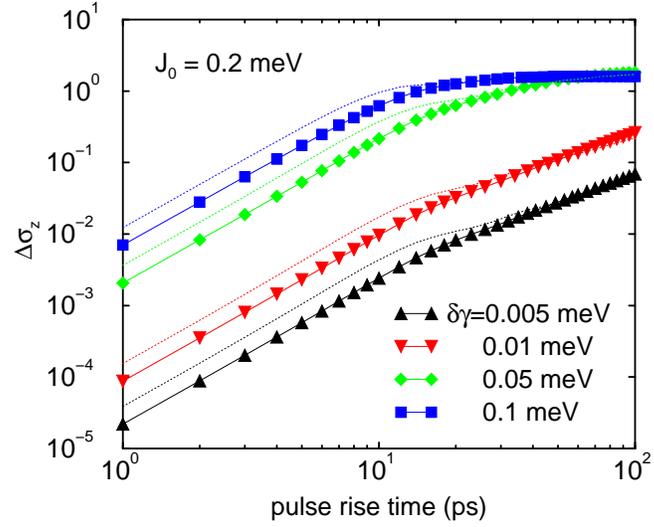}
\protect\caption[fidelity vs. pulse rise time]
{\sloppy{
Single pulse swap infidelity $\Delta \sigma_z$ as a function of pulse rise
time at a fixed exchange coupling $J$ and four different Zeeman
inhomogeneities.  In this figure and the two following ones, the dashed lines
without symbols represent data for trapezoidal pulse shape (linear
rise/fall), while the solid lines with symbols are for a sinusoidal pulse
rise/fall.
}}
\label{fig:rise}
\end{figure}

\begin{figure}
\includegraphics[width=3.6in]
{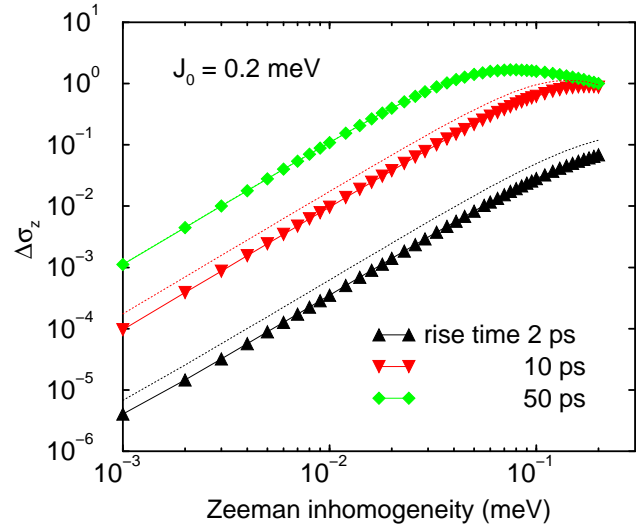}
\protect\caption[fidelity vs. inhomogeneity]
{\sloppy{
Single pulse swap infidelity $\Delta \sigma_z$ as a function of Zeeman
inhomogeneity at a fixed exchange coupling $J$ and three different pulse rise
times.
}}
\label{fig:inh}
\end{figure}

\begin{figure}
\includegraphics[width=3.6in]
{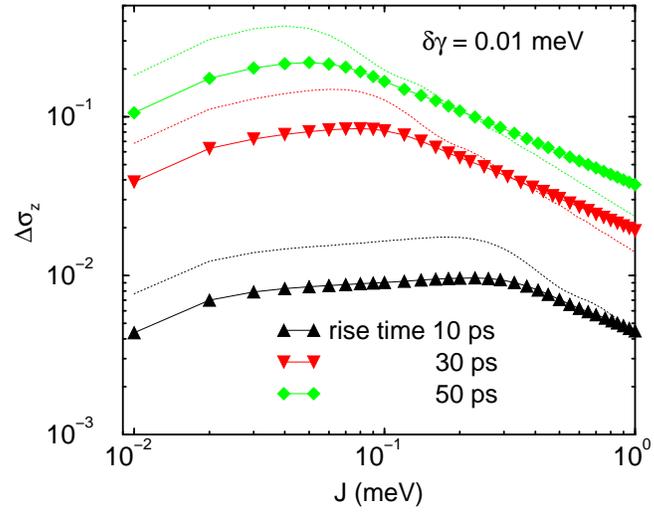}
\protect\caption[fidelity vs. J]
{\sloppy{
Single pulse swap infidelity $\Delta \sigma_z$ as a function of the exchange
coupling $J$ at a fixed Zeeman inhomogeneity and three different pulse rise
times.
}}
\label{fig:J}
\end{figure}

In Figures \ref{fig:rise}, \ref{fig:inh}, and \ref{fig:J}, we show
how the efficiency of an exchange pulse
in performing rotations varies with the exchange coupling $J$, Zeeman
inhomogeneity $\delta$, and the pulse rise/fall time $\tau_r$.  In these
figures, the filled and unfilled symbols correspond to two different pulse
shapes: sinusoidal and linear rise/fall, respectively.  For the sinusoidal
rise, $J=J_M sin^2(\pi t/2\tau_r)$, while for the linear rise, $J=J_M
t/\tau_r$, where $J_M$ is the maximal exchange couping during the pulse.

In Fig.~\ref{fig:rise}
we plot the effective infidelity $\Delta \sigma_z$ as a function
of the pulse rise/fall time when the maximum exchange is set at $0.2$meV.
There are four sets (each with two pulse shapes) of data shown in the
figure corresponding to four different values of Zeeman inhomogeneity, as
indicated in the legend.  At small pulse rise/fall time $\tau$, the
infidelity $\Delta \sigma_z$ grows approximately as a quadratic function of
$\tau$ (below $\sim 10$ps) for both pulse shape.  At longer $\tau$ the
growth saturates since the maximum value of $\Delta \sigma_z$ is 2.
In Fig.~\ref{fig:inh}
we plot the effective infidelity $\Delta \sigma_z$ as a function
of the Zeeman inhomogeneity at a fixed exchange coupling $J=0.2$meV
and three different pulse rise times.  Similar to Fig.~\ref{fig:rise},
the infidelity
grows quadratically (for both pulse shapes) with inhomogeneity, then
saturates when the inhomogeneity is in the same order of magnitude as the
exchange coupling.  The quadratic behavior in both 
Fig.~\ref{fig:rise} and Fig.~\ref{fig:inh} might
be understood as the result of small parameter Taylor expansions.
In Fig.~\ref{fig:J}, the infidelity $\Delta \sigma_z$ is plotted as a
function of the
maximum exchange coupling $J$ at a fixed inhomogeneity $\delta\gamma =
0.01$meV and three different pulse rise times.  Here for the two different
pulse shapes the infidelity decreases with different power law as the
exchange coupling $J$ increases, essentially because the larger exchange
coupling leads to an enhanced role played by the time-dependent function of
the pulse shape.

\begin{figure}
\includegraphics[width=3.6in]
{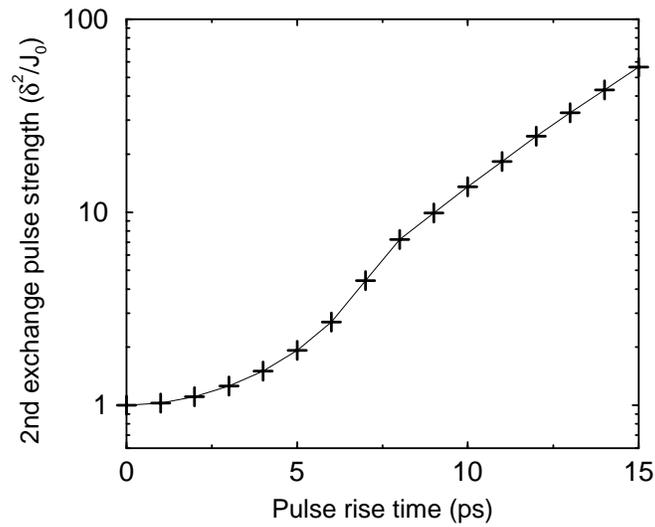}
\protect\caption[2nd pulse strength vs. pulse rise time]
{\sloppy{
Strength of the second exchange pulse in a complete swap pulse sequence as
a function of the pulse rise time.  The first exchange pulse strength is
set at 0.2 meV while Zeeman inhomogeneity is 0.1 meV.
}}
\label{fig:2ndJstrength}
\end{figure}

It is clear from Figs.~\ref{fig:rise} to \ref{fig:J} that
imperfect exchange pulses (or
non-square pulses) in the presence of Zeeman inhomogeneity reduce
the effectiveness of the exchange interaction in performing swap operation.
The key question now is whether one can still perform swap operations when
square pulses are not available.  The answer is affirmative.  In analogy to
the square pulse case shown in Fig.~\ref{fig:pulses}, one can turn on the 
most efficient exchange pulse as shown above, then let the system freely 
evolve (in the presence of Zeeman inhomogeneity) an optimal period of time,
then turn on an exchange pulse again.  Now each of the pulses is generally
not an exact $\pi$ pulse, but have to be calculated numerically to last for
the most efficient duration.  Similar to the square pulse case, such pulse
sequence might have to be used more than once if the Zeeman inhomogeneity is
too large while the 
exchange coupling strength is weak.  However, in many realistic situations,
some numerical corrections to the pulse sequence shown in
Fig.~\ref{fig:pulses} should be sufficient to produce a precise swap 
operation.  In Table~\ref{tab:pulse_seq} we summarize three different
situations.
\begin{table}[h]
\begin{tabular}{|l|c|c|c|} 
\hline
Pulse sequence & 1 & 2 & 3 \\ \hline
Zeeman inhomogeneity (meV) & 0.1 & 0.1 & 0.01 \\ \hline
pulse rise time (ps) & 5 & 10 & 5 \\ \hline
1st exchange coupling (meV) & 0.2 & 0.2 & 0.2 \\ \hline
2nd exchange coupling (meV) & 0.0965 & 0.677 & 0.000605 \\ \hline
corrections to the 2nd pulse (meV) & +0.0465 & +0.672 & +0.000105 \\ \hline
fidelity & 0.999998 & 0.999994 & 0.9999999 \\ \hline
\end{tabular}
\caption[]{Numerically obtained pulse sequences for various exchange
pulses to achieve a complete swap.  Notice that in each of the cases
the strength of the second exchange pulse increases, by 93\%, 12.5 times,
and 20\%, respectively.}
\label{tab:pulse_seq}
\end{table}
Here we can see that longer pulse rise time leads to significantly higher
requirement for large exchange interaction.  In the case of the first and
second pulse sequences, if the pulses are square, the magnitude of the second
exchange pulse needs to be $\delta^2/J = 0.05$ meV.  When the pulse rise time
is 5 ps, this magnitude nearly doubles; while a 10 ps rise time leads to a
large exchange coupling of 0.677 meV, more than ten times bigger than the
square pulse case.  If the maximum exchange is below 0.5 meV, a swap would
require several pulses to perform (as in Eq.~(\ref{eq:swap_multi})), so that
the operation becomes less efficient.  To further illustrate this point, we
plot the strength of the second exchange pulse $J^{\prime}$ in the 3-pulse
swap sequence of Eq.~(\ref{eq:swap}) as a function of the pulse rise time in
Fig.~\ref{fig:2ndJstrength}.  The unit for $J^{\prime}$ is its value in the
square pulse case: $\delta^2/J_0$ where $\delta$ is the Zeeman inhomogeneity
and $J_0$ is the strength of the first exchange pulse (with the same pulse
rise time for simplicity).  It is quite clear from
Fig.~\ref{fig:2ndJstrength} that the required strength for the second
exchange pulse increases exponentially with the pulse rise time.  For
example, at 15 ps pulse rise time, the required $J^{\prime}$ is above 2.8 meV
(as compared to 0.05 meV in the square pulse case), beyond the range
available in the current state-of-the-art double dots.  Multi-pulse sequences
as in Eq.~(\ref{eq:swap_multi}) would have to be invoked in such scenarios.

In summary, we have assessed the effects of qubit inhomogeneity on the
swap operations in a quantum dot quantum computer.  We have shown that 
the exchange Hamiltonian becomes less efficient in performing swap
when qubit inhomogeneity is present and the exchange is not turned on and off
in the ideal square pulse shape.  We have demonstrated ways to perform
complete swap, at the expense of longer pulse sequences and
numerically-searched pulse parameters \cite{zerofield}.  In exchange-based
quantum computing schemes, entangling operations such as controlled-NOT are
constructed from the square root of swap operation.  With the extra
complexity in the swap operation in the currently studied situation, it is
quite natural to expect more complexity in controlled-NOT operations as well. 
Furthermore, since our calculation is performed in an effective two-level
system, our results are applicable to single qubit operations, too.  In
essence, if $|0\rangle$ and $|1\rangle$ states are split energetically,
attempts to perform direct rotations around axes other than $z$ will be
hindered if that interaction is turned on and off gradually instead of in a
square pulse profile.  Indeed, the effects of such non-ideal pulses have been
explored in solid-state systems like Cooper pair boxes \cite{Choi,Oh,Gate}
and charge oscillations in double quantum dots \cite{Fuji1,Fuji2,Fuji3}, and
will surely be encountered even more in the future experimental studies of
various solid-state quantum computing architectures.

This work is supported by ARDA and LPS.


\begin{thebibliography}{99}
\bibitem{note} The study of this effective spin-$1/2$ system has been done
in the context of Ref.~\cite{Levy} in terms of square pulses
\cite{Geometric}, while we in this paper focus on the geometric explanation
of swap and the effects of the nonsquare pulses.
\bibitem{zerofield} Alternative approaches (such as turning off the applied
magnetic field), which go beyond the scope of the current study, can be
designed to overcome the problem caused by inhomogeneous Zeeman splittings.
\end{thebibliography}
\end{document}